\documentclass[showpacs,english,twocolumn,prl]{revtex4}
\usepackage[T1]{fontenc}
\usepackage[latin1]{inputenc}
\usepackage{graphicx}
\makeatletter

\usepackage{babel}
\makeatother
\begin{document}

\title{Giant fluctuations of superconducting order parameter in ferromagnet/superconductor single electron transistors}

\author{J. Johansson, V. Korenivski, D. B. Haviland}
\affiliation{Nanostructure Physics, Royal Institute of Technology, AlbaNova University Center, 106 91 Stockholm, Sweden }

\author{Arne Brataas}
\affiliation{Department of Physics, Norwegian University of Science and Technology, N-7491 Trondheim, Norway}

\begin{abstract}
Spin dependent transport in a ferromagnet/superconductor/ferromagnet single electron transistor is studied theoretically with spin accumulation, spin relaxation, gap suppression, and charging effects taken into account. A strong dependence of the gap on the magnetic state of the outer electrodes is found, which gives rise to a negative magneto-resistance of up to 100 \%. We predict that fluctuations of the spin accumulation due to tunneling of quasi-particles can play such an important role as to cause the island to fluctuate between the superconducting
and normal state. 
\end{abstract}


\maketitle
Electronic devices, which exploit the spin as well as charge of the electron have attracted much interest over the last several years. This interest in ``spintronics'' is stimulated by expected industrial applications, such as magnetic tunnel junctions based sensors, magnetic random access memory, as well as basic physics, offering new ways to manipulate quantum states using these two fundamental properties of the electron.

In nanoscale magnetic devices, such as single electron transistors (SET's), the presence of Coulomb blockade makes the charge transport sensitive to a very few electron states, which results in a number of pronounced effects in magneto-conductance (see review \onlinecite{Martinek:jap:review} and refs. therein). The sensitivity to spin imbalance in the SET should be greatly enhanced if the island is made superconducting, since the dependence of the gap on the number of accumulated spins is highly non-linear. Takahashi \emph{et al.} \cite{Takahashi} analysed a double tunnel barrier structure with ferromagnetic outer layers and superconducting central layer (F/S/F) without charging effects. They found that the anti-parallel alignment of the ferromagnetic films can produce a non-equilibrium spin distribution in the central layer, which suppresses the superconducting gap and thereby strongly influences the transport. Immamura \emph{et al.} \cite{Imamura_FSF_parity} considered the parity effect in a ferromagnetic SET with a superconducting island (F/S/F SET) in the limiting case of charging energy larger than the superconducting gap, low voltage, and zero temperature. They found that the spin accumulation is zero in the limit considered, i.e., the superconducting gap of the island is unaffected. Magnetoresistance (MR) can exist due to the polarized spin of the single excess electron but is reduced in magnitude from the Julliere value by spin relaxation. 

In this letter we develop a comprehensive model of the F/S/F SET by taking into account spin accumulation, spin relaxation, gap suppression, charging effects as well as conduction fluctuations in experimentally realistic structures, at finite temperatures. We find that spin accumulation in the F/S/F SET suppresses the gap, which strongly affects the transport properties and results in a greatly enhanced MR. Progressively smaller spin accumulation is required to suppress the gap as the size of the island is reduced, making it feasible to electrically detect just a few non-equilibrium spins. We find that fluctuations of the spin accumulation can play such an important role as to cause giant fluctuations of the gap between the full value at a given temperature and zero (between the superconducting and normal state \emph{at fixed bias and gate voltages}), with corresponding large fluctuations in the conductance of the device. Even in the parallel magnetic state of the device, where the average spin accumulation is zero, fluctuations of spin accumulation can lead to a significant gap suppression. The transistor exhibits a practically full \emph{gate-controlled} MR effect (MR$\sim1$, corresponding to $\sim1000$\% in typical GMR notations). We compare the results of the model with the recent, rather divergent experimental findings on ferromagnetic SET's \cite{Chen,Johansson,Johansson_comment}.

The model geometry is shown in Fig.1(a). Two ferromagnetic electrodes are connected to a superconducting island through tunnel junctions. The tunnel resistance is assumed to be much larger than the Quantum resistance, $R_{i}\gg R_{Q} = h/e^2 \simeq 26\ \text{k}\Omega$. We assume collinear magnetization of the two electrodes, the magnetization of the left electrode can be either parallel (\emph{P}) or anti-parallel (\emph{AP}) to the magnetization of the right electrode. The resistance of the tunnel junctions is spin dependent: $R_{i}^{\uparrow}=2R_{i}/(1+P_{i})$, $R_{i}^{\downarrow}=2R_{i}/(1-P_{i})$ (where $i=L$ (left) or $R$ (right) denotes the junction). Here the polarization is of the same sign ($P_{L}P_{R}>0$) for the parallel alignment. Letting $P_{L}\rightarrow-P_{L}$ (or $P_{R}\rightarrow-P_{R}$) for reversed magnetization of the left (or right) electrode, we obtain the resistances in the anti-parallel alignment.

The spin relaxation time can be comparable to or greater than the average time between tunnel events, so that a non-equilibrium spin distribution can accumulate on the island, resulting in a shift in the chemical potential. The shift in chemical potential is schematically shown in Fig.1(b). We consider the regime where the energy relaxation time is much shorter than the average time between two successive tunnel events (no gap suppression is expected in the opposite, elastic regime; for details see \onlinecite{Tserkovnyak:FSF}). The electrons are then instantaneously thermalized after they have tunneled and are described by the Fermi distribution function. This shift in chemical potential, $\Delta \mu^\sigma$, is related to the deviation in the number of spins from the equilibrium value, $n^\sigma$, for each spin-band, $\sigma = \uparrow,\ \downarrow$: 
\begin{equation}
n^{\sigma}=\rho V\int_{-\infty}^{\infty}dE\,N(E)\left[f(E-\Delta\mu^{\sigma})-f(E)\right],\label{eq:Dmu}
\end{equation}
where $N(E)=\Theta(E-\Delta)\frac{|E|}{\sqrt{E^{2}-\Delta^{2}}}$  is the BCS density of states, $\rho$ is the density of states at the Fermi energy (per spin) in the normal state, $f(E)$ is the Fermi distribution function, $V$ is the island volume. The shift in the chemical potential on the superconducting island influences the gap through the BCS gap equation \cite{Tinkham}
\begin{equation}
\ln\left(\frac{\Delta_{0}}{\Delta}\right)=\int_{0}^{\hbar\omega_{D}}\frac{dE}{\xi}\left[f(\xi+\Delta\mu^{\uparrow})+f(\xi+\Delta\mu^{\downarrow})\right],\label{eq:BCS}\end{equation}
 where $\xi=\sqrt{E^{2}+\Delta^{2}}$ and $\omega_{D}$ is the Debye frequency. For arbitrary spin distribution on the island, (arbitrary $n^{\uparrow}$ and $n^{\downarrow}$), solving equations (\ref{eq:Dmu}) and (\ref{eq:BCS}) self-consistently yields the shift in the chemical potential, $\Delta\mu^{\sigma}$ , and the gap, $\Delta$. In Fig.2 the chemical potential and the gap as a function of the spin imbalance, $s\equiv n^{\uparrow}-n^{\downarrow}$, are shown for the case $n^{\uparrow}=-n^{\downarrow}$. When the spin imbalance increases, there will be a critical value of $s$ where the gap will discontinuously drop to zero and the chemical potential will drop to a lower value. After the complete suppression of the gap the chemical potential will increase linearly with the slope of $1/\rho V$ as in the normal conducting case. Reducing the spin imbalance will restore the gap at the same critical value of $s$. Note that the gap and chemical potential do not only depend on $n^{\uparrow}$ and $n^{\downarrow}$ through the combination $s=n^{\uparrow}-n^{\downarrow}$, the total spin accumulation. Different $n^{\uparrow}$ and $n^{\downarrow}$ having the same $s$ can produce different $\Delta$ and $\Delta\mu^{\sigma}$.

The transport properties are determined by the tunneling rates according to the ``orthodox theory''~\cite{SCT}. The tunneling rates depend on the charge of the island through $n=n^{\uparrow}+n^{\downarrow}$, as well as $n^{\uparrow}$ and $n^{\downarrow}$ separately, through the dependence of $\Delta$ and $\Delta\mu^{\sigma}$ on $n^{\uparrow}$ and $n^{\downarrow}$. The tunneling rates are spin dependent also through the spin dependent resistances: \begin{widetext}
\begin{eqnarray}
\Gamma_i^{\text{on},\sigma}(n) &=& \frac{1}{e^2R_i^{\sigma}}\int_{-\infty}^{\infty}dE N(E) f(E - \delta E_i^{\text{on}}(n))\left[1-f(E + \Delta \mu^{\sigma})\right] \nonumber \\
\Gamma_i^{\text{off},\sigma}(n) &=& \frac{1}{e^2R_i^{\sigma}}\int_{-\infty}^{\infty}dE N(E)f(E - \Delta \mu^{\sigma})\left[1-f(E  - \delta E_i^{\text{off}}(n))\right],\nonumber\end{eqnarray}
\end{widetext} where $\delta E_{i}^{\text{{on/off}}}(n)$ is the change in the charging energy for tunneling one electron on or off the island through the left or right junction:\[ 
\delta E_{i}^{\text{on/off}}(n)=\frac{e^{2}}{C_{\Sigma}}\left[\pm\left(n-n_{g}+\frac{V_{i}}{e}\frac{C_{L}C_{R}}{C_{i}}\right)+\frac{1}{2}\right]
\]
Here $C_{\Sigma}=C_{L}+C_{R}+C_{g}$ is the total capacitance of the island and $n_{g}=C_{g}V_{g}/e$ is the normalized gate charge. Tunneling that increases the the energy have exponentially small rates (zero at $T=0$), giving the characteristic Coulomb blockade below a threshold voltage.

In what follows we choose the parameter space to represent the typical nano-lithographically fabricated SET structures. We therefore consider SET's with island dimensions in the $10^{5}-10^{6}$ nm$^{3}$ range, such that up to a hundred states will participate in the transport. The relatively large number of states involved makes the Monte Carlo method \cite{Likharev} more suitable than the master equation approach. The intertwined dependence of the gap, the chemical potential and the spin accumulation can be included in a straight forward way with the Monte Carlo method. Moreover, the fundamental stochastic nature of tunneling and the resultant fluctuations in the spin transport through the structure are well modeled by the Monte Carlo procedure. 

The state of the system is determined by $n^{\uparrow}$ and $n^{\downarrow}$, from which the gap and shift in chemical potential are calculated using Eqs. (\ref{eq:Dmu}) and (\ref{eq:BCS}). After obtaining $\Delta$ and $\Delta\mu^{\sigma}$, the eight possible tunneling rates are computed, which increase or decrease $n^{\uparrow}$ or $n^{\downarrow}$ by one. The interval zero to one is then divided into eight segments with widths proportional to the tunneling rates and a random number in the interval {[}0,1{]} is generated to choose one of the eight tunnel events. The probability to preserve the present configuration decays as $P=\exp[-(t-t_{0})\Gamma_{\Sigma}]$, where $t_{0}$ is the time of the previous tunnel event and $\Gamma_{\Sigma}$ is the sum of the eight tunneling rates. A random number in the interval {[}0,1{]} gives the time between successive tunnel events by inverting the equation for the probability. The spin flip process is then accounted for in a phenomenological way by letting $s\rightarrow s\exp[-(t-t_{0})/\tau_{\text{{sf}}}]$, where $\tau_{\text{{sf}}}$ is the spin flip time. This procedure is repeated, keeping track of all quantities, until good statistics is obtained.

The dynamic effect of gap suppression is shown in Fig.3, where the  spin accumulation and superconducting gap are plotted versus time at fixed bias and gate voltages. As soon as the spin imbalance reaches the critical value ($\sim18$ for these device parameters) the gap drops to zero and the island goes normal. At the critical bias, fluctuations of only a few polarized spins on the island cause giant fluctuations of the order parameter from the full value at the given temperature to zero. This effect should be directly observable in experiments employing time resolved (<1$\mu$s scale) conductance measurements. Typical current-voltage (I-V) measurements (>1 ms time constant) will record the gap, which is a time average of the dynamic gap shown in Fig.3. Fig.4 shows this mean value of the superconducting gap, $\langle\Delta\rangle$, as well as its standard deviation, $\delta\Delta=[\langle\Delta^{2}\rangle-\langle\Delta\rangle^{2}]^{1/2}$, as a function of bias voltage. The fluctuations result in a smooth rather than discontinuous transition from the superconducting to the normal state as the spin accumulation is increased in the \emph{AP} state by increasing the bias. The fluctuations peak at the critical bias in the \emph{AP} state (0.42 mV in this case). Contrary to what would be intuitively expected, fluctuations of the spin accumulation on the island in the \emph{P} state can be very large. The fluctuations increase with bias voltage and becomes as large as the critical spin accumulation giving a significant gap suppression. This unexpected effect is a consequence of the stochastic dynamics of \emph{spin}-dependent transport through the device (charge fluctuation is suppressed by the charging energy) and the highly non-linear dependence of the gap on spin accumulation, since the time averaged spin accumulation in the parallel state is zero.

The suppression of the superconducting gap by spin imbalance in the \emph{AP} state results in fluctuations of the conductance of the device, which are greater in magnitude as the conductance itself. Fig.5a shows the time averaged I-V characteristics of the ferromagnetic SET in both the \emph{AP} and \emph{P} states. At the critical bias, the gap suppression leads to a 3-fold increase in current as a function of the magnetic state of the device, which results in a correspondingly large negative magnetoresistance. An even stronger MR effect, MR>0.8 is obtained  at the critical bias by applying a gate voltage as shown in Fig.5b. The island size chosen for the illustration above is representative of typical experimental implementations today, but not the smallest fabricated to date. For smaller superconducting islands, progressively fewer spins are needed to produce the full effect of gap suppression, with MR reaching $\approx1$. Increasing the spin relaxation time or reducing thermal excitations by going to lower temperatures enhances the effect. 

To estimate the parameters of a realistic experimental realisation  of a F/S/F SET, assume a SET fabricated with Cobalt leads having a spin polarization of 0.35, Aluminum as the superconducting island with $\Delta=200\,\,\mu$eV, $\rho=10^{28}$ {[}m$^{3}$eV{]}$^{-1}$ and a volume of $15\times70\times200$ nm$^{3}$, so that $\rho V=2.1\cdot10^{6}$ {[}eV{]}$^{-1}$. The spin relaxation time of quasi-particles in superconducting \emph{Al} has not been measured, but for normal Aluminum at 4.2 K spin relaxation times between 10 ns \cite{Johnson} and recently 0.1 ns \cite{Jedema} have been reported. Assuming a spin relaxation time of $\tau_{\text{{sf}}}\approx$ 1 ns, $T=$0.05 K, $R=30$ k$\Omega$, and $E_{\text{{c}}}=100$ $\mu$eV this sample would have the transport characteristics shown in Figs.2-5. If the spin relaxation time was shorter, the island volume would have to be smaller to achieve the same effect. Increased $\tau_{sf}$ in the superconducting state has been predicted theoretically \cite{Yamashita}, which would lead to stronger gap suppression and higher MR.

When we compare our theoretical results with the recent experiments on F/S/F SET's \cite{Chen,Johansson}, we draw the conclusion that the reported MR is unlikely due to gap suppression by spin accumulation. The volume of the islands studied was rather large in both experiments ($\rho V\approx10^{7}$ {[}eV{]}$^{-1}$) and the temperature relatively high (T $\approx$ 0.25 K). The combined effect of large volume and high temperature would give insignificant gap suppression in the voltage interval where large MR was reported. There are several other points where our simulations do not agree with the interpretation of experiments reported. In both experiments the maximum MR was observed at or below the threshold voltage, $eV_{\text{{th}}}\le2\Delta+E_{\text{{c}}}$. In our calculations we always find $V_{\text{{th}}}$ to be the same in the two states, and the maximum MR occurs above $V_{\text{{th}}}$. This arises form the fact that the spin accumulation is directly proportional to the current, so there is no mechanism for spin accumulation below the threshold voltage, where no current flows. Further, the extracted gap in \cite{Chen} did not go to zero monotonically with increasing voltage but had a maximum at an intermediate voltage, contrary to what we predict. The temperature dependence of the MR measured in \cite{Johansson}, with the peak of the MR occurring at the same bias voltage, is in contrast with our calculations where the MR peak shifts to higher bias with increasing temperature (not shown). More experiments are needed to resolve these controversies.

\begin{acknowledgments}
J. Johansson gratefully acknowledge support from the Swedish SSF Spintronics program. 
\end{acknowledgments}

\bibliographystyle{apsrev}

\newpage
\begin{figure}
  \includegraphics[width=0.95\linewidth]{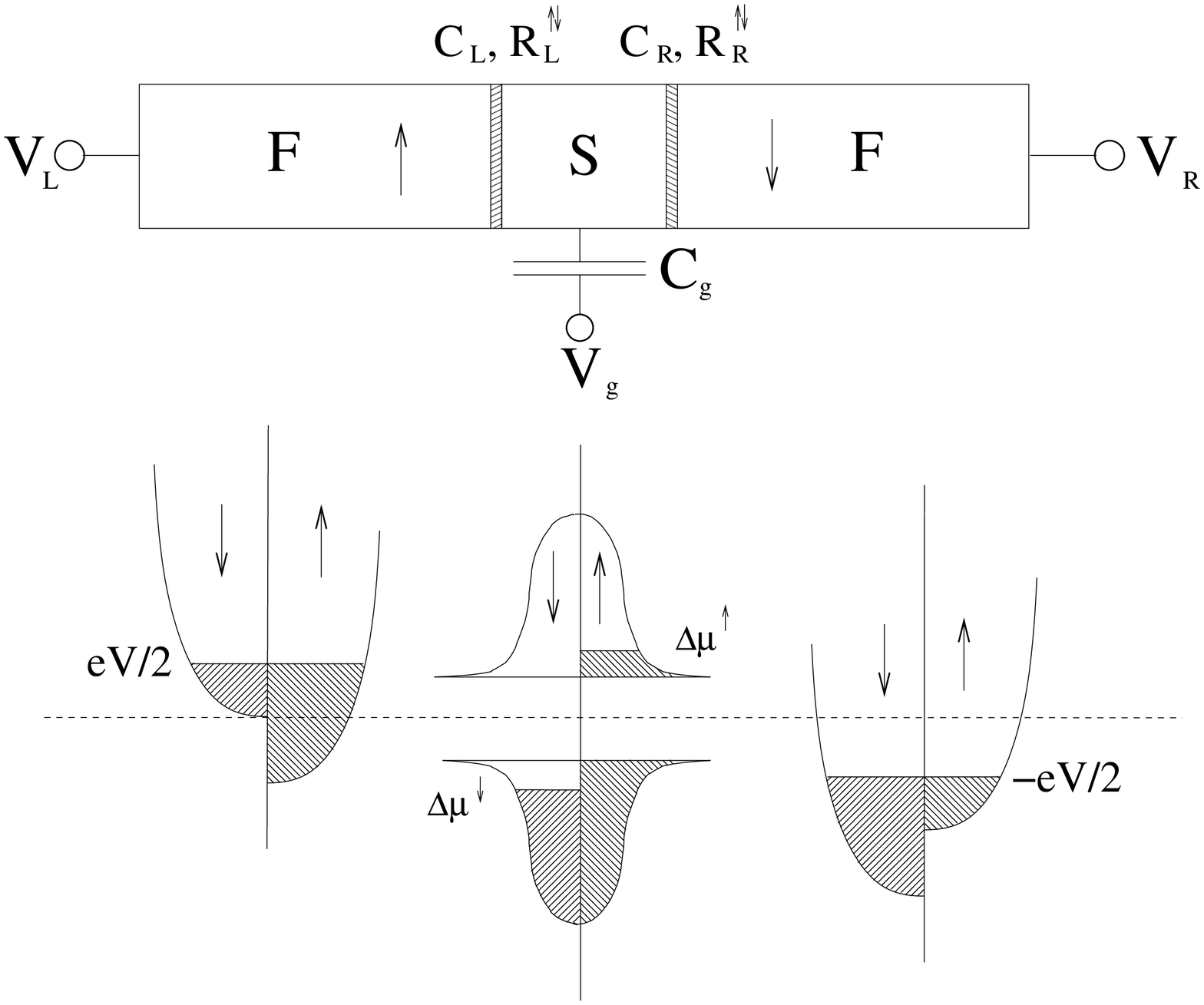}
  \caption{\label{fig:sample}a) The single electron transistor consisting of two ferromagnet (F) electrodes and a superconducting (S) island. b) Schematic density of states for the ferromagnet (left and right) and superconductor (middle) for the anti-parallel alignment of the electrodes, when $V_{L}=V/2$ and $V_{R}=-V/2$.}
\end{figure}

\begin{figure}
  \includegraphics[width=0.95\linewidth]{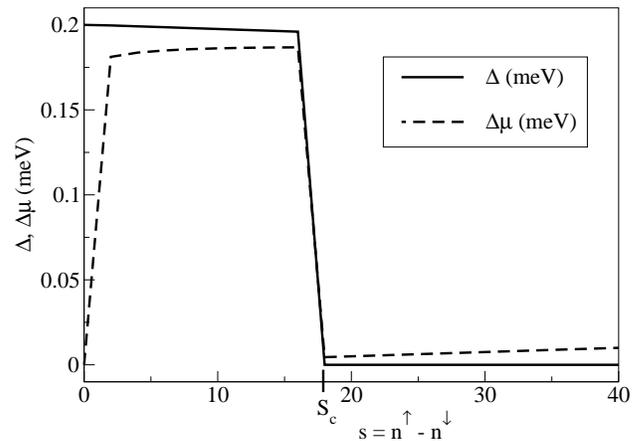}
  \caption{\label{fig:D_and_DE}The self-consistent solutions of Eq. (\ref{eq:Dmu}) and (\ref{eq:BCS}). The gap, $\Delta$, and chemical potential, $\Delta\mu$, as a function of the spin imbalance for $\rho V=2\cdot10^{6}$ {[}eV{]}$^{-1}$ and $T=0.05$ K. The critical spin accumulation, $S_{\text{{c}}}$, is shown.} 
\end{figure}

\begin{figure}
  \includegraphics[width=0.95\linewidth]{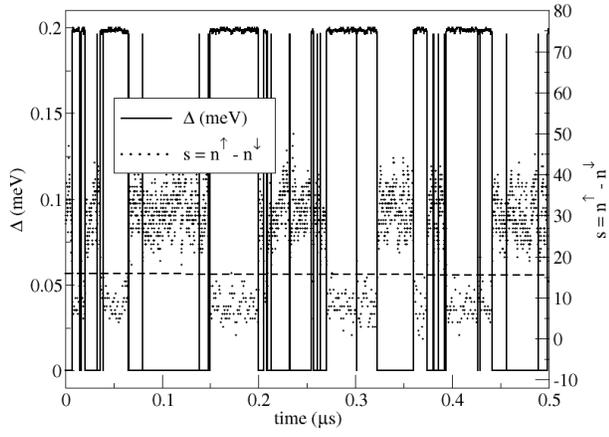}
  \caption{\label{fig:fluct}Time trace of the gap (solid line) and spin accumulation (dots) at $S_{\text{{c}}}$ in Fig.2, corresponding to $V=$0.42 mV and $V_{g}=$0 mV. The device parameters are $R_{L}=R_{R}=30$ k$\Omega$, $C_{L}=C_{R}=0.8$ fF, $P_{L}=P_{R}=0.35$, $\rho V=2\cdot10^{6}$ {[}eV{]}$^{-1}$, $T=0.05$ K, $\tau_{sf}=1$ ns. $S_{\text{{c}}}$ is indicated with a dashed horizontal line.}
\end{figure}

\begin{figure} 
  \includegraphics[width=0.95\linewidth]{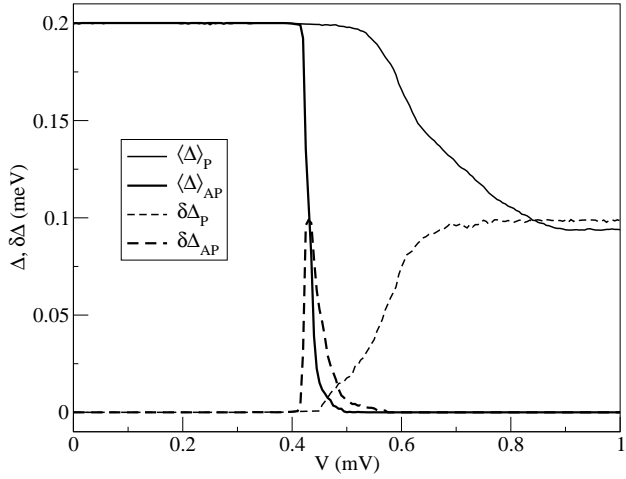}
  \caption{\label{fig:MR_vs_T}The time averaged value (solid) and standard deviation (dashed) of the superconducting gap for the \emph{P} and \emph{AP} states of the device in Fig.3, $V_{g}C_{g}/e=0.5$. }
\end{figure}

\begin{figure} 
  \includegraphics[width=0.95\linewidth]{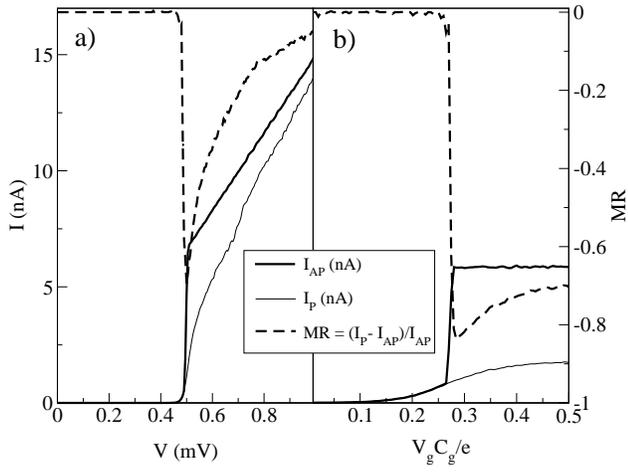}
  \caption{\label{fig:IV} a) \emph{I-V} characteristics in the \emph{P} and \emph{AP} state and MR of F/S/F SET with same parameters as in Fig.3 and $V_{g}=0$; b) the gate dependence at V=0.45 mV.}
\end{figure}

\end{document}